\documentclass[
reprint,
prl,
superscriptaddress,
showpacs,
amsmath,
amssymb,
aps,
]{revtex4-2}

\newcommand{\rv}{{\bf r}}
\newcommand{\Rv}{{\bf R}}
\newcommand{\Xv}{{\bf X}}
\newcommand{\kv}{{\bf k}}
\newcommand{\iv}{{\bf i}}
\newcommand{\jv}{{\bf j}}
\usepackage{lmodern}
\usepackage{lineno}
\usepackage{mathtools}
\usepackage{color,soul}
\usepackage{graphicx}
\usepackage{dcolumn}
\usepackage{bm}
\usepackage{feynmp-auto}
\usepackage{amssymb,amsfonts,amsmath}
\usepackage{threeparttable}
\usepackage{textcomp}
\usepackage{float}
\usepackage{wrapfig}
\usepackage{verbatim}
\usepackage{accents}
\usepackage{fancyhdr}
\usepackage{euscript}
\usepackage{circledtext}
\usepackage{blkarray}
\usepackage{multirow}
\usepackage{tikz}
\usepackage{subfigure}
\usepackage{enumitem}
\usepackage{physics}
\usepackage[T1]{fontenc}
\usepackage{newtxtext,newtxmath}
\usetikzlibrary[topaths]
\usepackage[breaklinks=true, colorlinks, 
citecolor=blue,linkcolor=blue,urlcolor=blue]{hyperref}

\graphicspath{ {./figs/} } 

\begin{document}
\title{Orbital Magnetic Moment Dynamics and Hanle Magnetoresistance in Multilayered 2D Materials}

\author{Hao Sun}
\email{sun.hao@nus.edu.sg}
\affiliation{The Institute for Functional Intelligent Materials (I-FIM), National University of Singapore, 4 Science Drive 2, Singapore 117544}

\author{Giovanni Vignale}
\email{vgnl.g@nus.edu.sg}
\affiliation{The Institute for Functional Intelligent Materials (I-FIM), National University of Singapore, 4 Science Drive 2, Singapore 117544}
 

\begin{abstract}
The orbital Hall effect (OHE), resulting from non-trivial quantum geometry of 2D materials, has several potential advantages over the spin Hall effect (SHE), the latter being well known for its many applications in spintronics.  Like the spin Hall effect, the OHE occurs in nonmagnetic materials without stringent symmetry requirements, but unlike the SHE it does no rely on  relatively weak spin-orbit interaction.  In  2D materials, these advantages risk to be nullified by the difficulty of turning the orbital moment away from the out-of-plane direction. Multilayered 2D materials offer a way out of this difficulty because the fluctuating in-plane component of the orbital moment, due to motion of electrons between the layers,  can latch to a magnetic field.  To describe this effect we have derived a semi-phenomenological equation of motion for the density of orbital magnetic moment in  stacked 2D  materials subjected to a magnetic field.  Unlike the equations of motion for the spin, these equations produce a strongly anisotropic dynamics, which is governed by an inverse effective mass tensor for which we provide a fully microscopic expression.  As a first application, we combine our  equation of motion with phenomenological drift-diffusion equations to obtain a theory of orbital Hanle magnetoresistance in multilayered 2D materials.

\end{abstract}

\maketitle

\textit{Introduction.---}The orbital Hall effect (OHE) ~\cite{Bernevig2005, Bhowal2020, Canonico2020, Cysne2021, Cysne2022, Chen2024, Pezo2022, Chang1996, Go2018, Dongwook2020,Go_2021, Pezo2023, Salemi2022, Aryasetiawan2019,Atencia2024,Rhonald2024,Costa2023} is the generation of a transverse current of orbital magnetic moments (OMM) in response to an applied electric field.  Similar to the better established (and very useful) spin Hall effect (SHE)~\cite{Ruiz2022, Liu2018, Saburo2008, Sinova2004, Tanaka2008, Tanaka2010, Sinova2015, Hoffmann2013, Manchon2015, Ralph2008, Hirsch1999, Dyakonov1971, Saitoh2006, Valenzuela2006}, the OHE can occur in non-magnetic materials in both two and three dimensions without stringent symmetry requirements; for instance, inversion symmetry breaking is not required. However, unlike the SHE, the OHE does not rely on relatively weak spin-orbit interactions and can, therefore, be prominent in light materials. On a microscopic level, OMM arises from both the angular momentum of intra-atomic orbitals and the inter-atomic motion of itinerant electrons. Both components are captured in the ``modern theory'', which expresses the orbital moment of Bloch electrons as a geometric property of the Bloch wave functions~\cite{Thonhauser2005, Ceresoli2006, Shi2007}.

Experimentally, OHE is revealed by the observation of magnetic moment accumulations~\cite{Mak2018, Mak2019, Zhang2020, Marui2023, Lyalin2023, Ghoi2023} near the sample edges and, less directly, by nonlocal resistance measurements in which OHE and its inverse act together to produce the measured potential difference~\cite{Stephen2021,Arrighi2023}. A third and more subtle manifestation is the magnetoresistive effect resulting from the accumulation of OMMs at the edges. This effect occurs because nonuniform edge accumulations of OMMs generate electric currents parallel to the edge, thereby slightly reducing the resistance of the sample. This effect, which has been dubbed Hanle magnetoresistance (HMR) in the context of the spin Hall effect, was first predicted theoretically~\cite{Dyakonov1994} and subsequently verified experimentally in Ref.~\cite{Casanova2009,Velez2016}. Its orbital version has recently been observed in 3D Mn thin polycrystalline films~\cite{Sala2023}, but not in crystalline 2D materials. 

\begin{figure}[t]
    \includegraphics[width=0.40\textwidth]{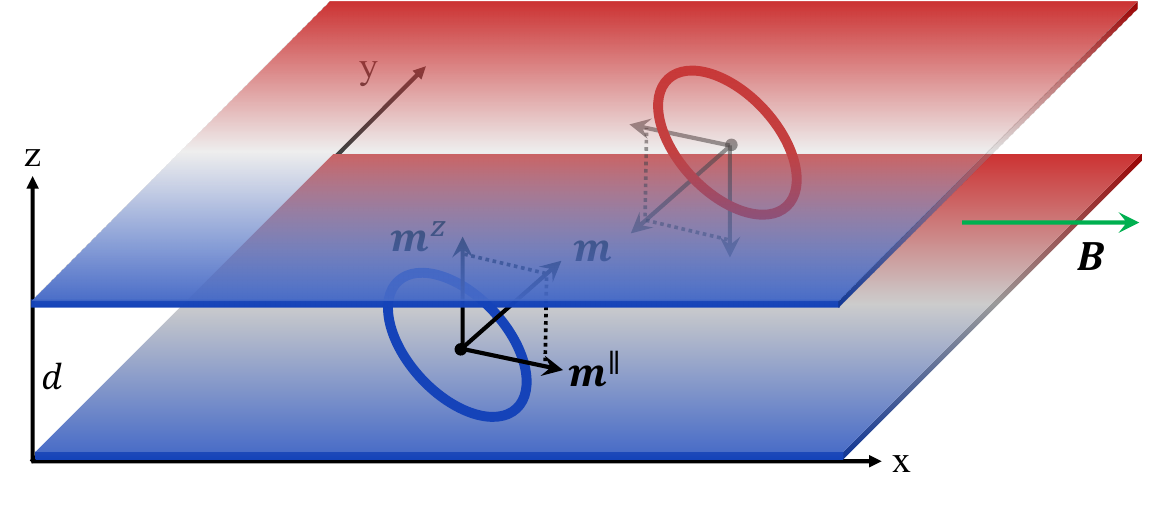}
    \caption{\label{fig1} Schematic plot of the in-plane orbital magnetic moment in a bilayer system. Its in-plane component $\mathbf{m}^{\shortparallel}$, generated by inter-layer motion, couples to the in-plane component of the magnetic field.}
\end{figure}

2D materials offer several distinct advantages for advancing OHE-based devices. First, their atomically thin profiles allow for strong confinement of electronic states, enhancing the sensitivity of charge response to external fields~\cite{Castro2009}. Second, 2D materials often exhibit high electronic mobility, even at room temperature~\cite{Morozov2008,Castro2009}, allowing efficient OHE-based transport without significant energy losses. Additionally, the  band structure of 2D materials can be tuned by strain, electric fields, or stacking configurations in multilayers~\cite{Guinea2010}.
However, these attractive features risk to be nullified by the difficulty of applying torques to control the magnitude and direction of the orbital moment accumulations~\cite{Cysne2023}. Unlike spin, which is easily controlled by magnetic fields, the OMM interacts strongly with the lattice, and its orientation is difficult to alter. 
At the extreme 2D limit, the OMM is locked to the out-of-plane direction, making such manipulation nearly impossible.
This poses the question: how can we effectively manipulate the OMM in 2D materials?

In this Letter, we show that {\it multilayered} 2D systems offer a promising solution for achieving precise, tunable control of orbital moments without the assistance of the spin-orbit interaction. The key enabling feature is the emergence of a fluctuating in-plane OMM arising from the looping motion of electrons between the layers. This in-plane OMM component provides a ``handle'' by which we can ``grab'' the OMM and reorient it away from the out-of-plane direction, as illustrated in Fig.~\ref{fig1}.

Our central result  is the equation of motion for the OMM density, denoted by $n^a_m$ -- a function of position and time:
\begin{equation}\label{EOM_Mac}
    \begin{aligned}[b]
        \frac{\partial }{\partial t}n^a_m=&-\nabla \cdot \mathbf{J}_m^a-\frac{n^a_m}{\tau_m}-\frac{e}{2}\epsilon^{abc}B_b X_{cd} n_m^d\,.
    \end{aligned}
\end{equation}
where $\tau_m$ is a phenomenological relaxation time arising from the combined action of the crystal field, impurity scattering processes, and intrinsic orbital moment torque, and $\mathbf{J}_m^a$ is the orbital current driven by the electric field. 
While $\tau_m$  closely corresponds to the spin relaxation time, the last term on the right hand side of Eq.~(\ref{EOM_Mac}) presents a major departure from spin dynamics  due to the appearance of the anisotropic inverse mass tensor $X_{cd}$ of the multi-layer system~\footnote{This effect could be disregarded in the spin case, being of higher order in the strength of spin-orbit coupling.}.  
The microscopic definition of $X_{cd}$ is
\begin{equation}\label{X_tensor}
    \begin{aligned}[b]
    X_{cd}=&\frac{1}{i\hbar}\langle[\hat{r}_c,\hat{v}_{d}]\rangle_F\,\\
    \equiv&\frac{1}{i\hbar}\sum_{n\mathbf{k}}\mel{n\mathbf{k}}{[\hat{r}_c,\hat{v}_{d}]}{n\mathbf{k}}f_{n\mathbf{k}},
    \end{aligned}
\end{equation}
where $\hat{r}_c$ and $\hat{v}_d$ are components of the position and velocity operators and $f_{n\mathbf{k}}$ is the Fermi-Dirac distribution function. This tensor is an equilibrium state property that should be calculated in the absence of electric and magnetic fields and can be shown to be symmetric under the interchange of indices $c,d$.  A derivation of Eqs.~(\ref{EOM_Mac}) and (\ref{X_tensor})  will be presented below, with further details in~\cite{supp}.

The  essential difference between the out-of-plane and the in-plane components of the inverse mass tensor is that the former is expressed entirely in terms of band-theoretical properties\cite{Resta_2018}, while the latter  involves interlayer displacements and velocities, which are not describable in terms of Bloch wave functions.  As a result, the dynamics of the OMM is strongly anisotropic, with precession frequencies around an in-plane axis being proportional to interlayer hopping amplitudes and typically much smaller than their out-of-plane counterparts.  As a first application, we combine the OMM dynamics with drift-diffusion equations for the current and the direct and inverse orbital Hall angles to obtain a general-purpose theory of orbital HMR. We illustrate the theory by calculating the HMR for a simple model of bilayer graphene subjected to crossed electric and in-plane magnetic fields and show that the HMR depends on various tunable parameters, such as the layer separation and the strength of the electric and magnetic fields. Comparing to the HMR recently observed in 3D polycrystalline films of Mn\cite{Sala2023}, we find that our model predicts a larger and more tunable HMR due to the higher mobility of the Dirac electrons compared to $d$-band electrons in Mn.

\textit{Microscopic derivation of the OMM dynamics.--} In this section, we present a brief derivation of Eq.~\eqref{EOM_Mac}. Starting from the OMM operator expression $\hat{m}^a=-\frac{e}{2}\epsilon^{abc}\hat{r}_b\star\hat{v}_c$, where $\hat A\star \hat B=\frac{1}{2}(\hat A \hat B+\hat B\hat A)$ is the symmetrized product of two operators, we define the OMM density operator as
\begin{equation}\label{}
    \begin{aligned}[b]
    \hat{n}^a_{m}(\mathbf{r})=\sum_{p}\hat{m}^a_p\star\delta(\mathbf{r}-\hat{\mathbf{r}}_p)\,,
    \end{aligned}
\end{equation}
where the subscript $p$ labels the single electron operators (i.e., $\hat{\mathbf{r}}_p$ is the position operator for the $p$-th electron).
The dynamics of OMM density is described by the Heisenberg equation of motion~\cite{supp}
\begin{equation}\label{HEQ}
    \begin{aligned}[b]
    \frac{\partial \hat{n}^a_{m}(\mathbf{r})}{\partial t}=\frac{1}{i\hbar}\left[\hat{n}^a_{m}(\mathbf{r}),\hat H_{em}\right],
    \end{aligned}
\end{equation}
where $\hat H_{em}=\hat H_0+\sum_p\left[e\mathbf{E}\cdot\hat{\mathbf{r}}_p+\mathbf{B}\cdot \hat{\mathbf{m}}_p\right]$ is the Hamiltonian (including electric and magnetic field terms) that drives  the system in the OHE. 
Using the fact that the particle density operator obeys the equation of motion $\partial_t \delta(\mathbf{r}-\hat{\mathbf{r}}_p)=-\nabla_{\mathbf{r}} \cdot \hat{\mathbf{v}}_p \star \delta(\mathbf{r}-\hat{\mathbf{r}}_p)$, where $\hat{\mathbf{v}}_p=\frac{1}{i\hbar}[\hat{\mathbf{r}}_p, \hat{H}_{em}]$ is the velocity operator, we rewrite  Eq.~(\ref{HEQ}) as
\begin{equation}\label{GCE}
    \begin{aligned}[b]
        \frac{\partial \hat{n}^a_{m}(\mathbf{r})}{\partial t}=-\nabla_{\mathbf{r}}\cdot\hat{\mathbf{J}}_{\mathbf{m}}^a+\sum_{p}\left(\partial_ t\hat{m}_p^a\right)\star\delta(\mathbf{r}-\hat{\mathbf{r}}_p).
    \end{aligned}
\end{equation}
where $\hat{\mathbf{J}}_{\mathbf{m}}=\sum_p\hat{\mathbf{m}}_p\star\hat{\mathbf{v}}_p\star\delta(\mathbf{r}-\hat{\mathbf{r}}_p)$ is the OMM current density.  The first term on the right-hand side, where $\partial_t\hat{\mathbf{m}}_p=i\hbar^{-1}[\hat{\mathbf{m}}_p,\hat{H}_{em}]$, is the total torque, which is responsible for the non-conservation of the OMM. 

Eq.~(\ref{GCE}) must be averaged over the non-equilibrium state driven by the electric field.
Setting the magnetic field to zero, the torque term arising from the non-commutativity of $\hat H_0$ and $\hat m_i^a$ can be phenomenologically described as follows:
\begin{equation}\label{e_d}
    \begin{aligned}[b]
        \left\langle\sum_p(\partial_t \hat{m}^a_p)\star\delta(\mathbf{r}-\hat{\mathbf{r}}_p)\right\rangle_{NE} =-\frac{{n}^a_{m}(\mathbf{r})}{\tau_{\mathbf{m}}}.
    \end{aligned}
\end{equation}
where $\tau_{\mathbf{m}}$ is the  relaxation time for the OMM.
Here ${n}^a_{m}(\mathbf{r}) = \langle \hat {n}^a_{m}(\mathbf{r})\rangle_{NE}$, where $\langle... \rangle_{NE}$ denotes the average in the non-equilibrium state induced by the electric field (the average torque in the equilibrium state is zero).  This coincides with the second term on the right-hand side of Eq.~(\ref{EOM_Mac}). 

Turning on the magnetic field,  we get the crucial torque term, which produces the magnetoresistance effect.  This is given by
\begin{equation}\label{Btorque}
    \begin{aligned}[b]
        \frac{1}{i\hbar}\left\langle \sum_p[\hat{\mathbf{m}}_p,\hat{\mathbf{m}}_p\cdot\mathbf{B}] \star\delta(\mathbf{r}-\hat{\mathbf{r}}_p) \right\rangle_{NE},
    \end{aligned}
\end{equation}
which is calculated with the help of the commutation relation
\begin{equation}\label{omm_alg}
    \begin{aligned}[b]
    [\hat{m}_p^a,\hat{m}_p^b]=\frac{e\hbar}{2i}\epsilon^{abc}X_{cd}\hat{m}_p^d.
    \end{aligned}
\end{equation}
with $X_{cd}$ defined in Eq.~(\ref{X_tensor}).
In order to obtain this, we have neglected the magnetic field dependence of the velocity operator, which arises from the orbital coupling $\hat{\mathbf{m}}\cdot\mathbf{B}$ and contributes only at higher order in $B$.
Employing Eq.~\eqref{omm_alg} and neglecting the high-order terms, the magnetic-driven torque can be written as 
\begin{equation}\label{m_d}
    \begin{aligned}[b]
        \sum_p(\partial_t \hat{m}^a_p)\star\delta(\mathbf{r}-\hat{\mathbf{r}}_p)\vert_B=-\frac{e}{2}\epsilon^{abc}B_bX_{cd}\hat{n}^d_{m}\,.
    \end{aligned}
\end{equation}
Taking the average of this in the non-equilibrium state, we recover the last term on the right-hand side of Eq.~(\ref{EOM_Mac}).

\begin{figure}[t]
    \includegraphics[width=0.48\textwidth]{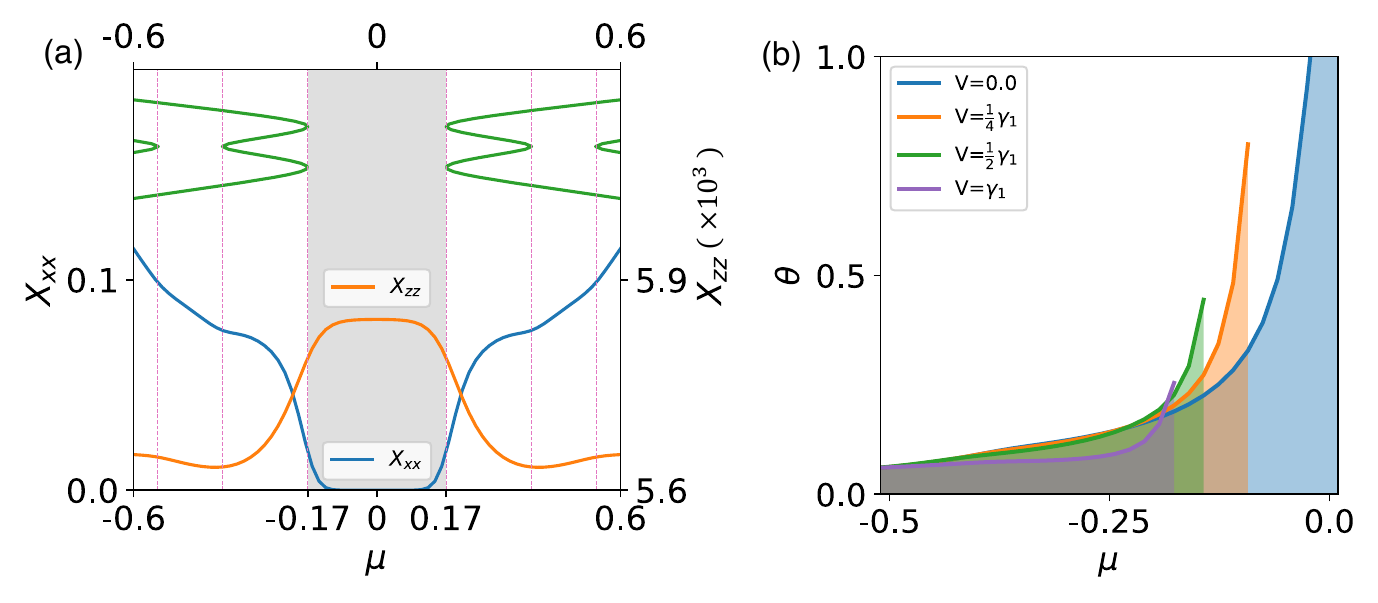}
    \caption{\label{fig3} (a) Principal values of the inverse effective mass tensor $X$ (The unit is $[m_e]^{-1}$, $m_e$ is the bare mass of electron) vs chemical potential in bilayer graphene (BLG) at room temperature (300K). $X_{xx}$ is plotted on the left $y$-axis and $X_{zz}$ on the right $y$-axis. Notice that the scale for  $X_{zz}$ has been multiplied by $10^3$. The shaded area indicates the insulating gap at a displacement field $V = \gamma_1$. For all calculations, we set $\gamma_0 = -3.16$ eV and $\gamma_1 = -0.38$ eV. The green lines above the $X$ curves show the band spectrum of BLG along the high-symmetry path $M-K-\Gamma$ centered at $K$ points. (b) The anisotropy ratio $\theta = \frac{X_{zz}}{X_{yy}}$ vs the displacement field $V$. Notice that the ratio is meaningless within the gap, where $X_{xx} = 0$.}
\end{figure}

\textit{ The physical meaning of the $X_{ab}$ tensor.--} 
We now consider more closely the inverse mass tensor $X_{ab}$ defined in Eq.~(\ref{X_tensor}). The general expression can be obtained for a general multi-layer Hamiltonian 
\begin{equation}\label{}
    \begin{aligned}[b]
        \hat{H}(\mathbf{k}) = &\sum_{l=1}^N \left(h(\mathbf{k})+V_lI\right) \hat{c}^{\dagger}_{\mathbf{k},l}\hat{c}_{\mathbf{k},l}\\
        &+\sum_{l=1}^{N-1}\left(\Gamma(\mathbf{k})\hat{c}^{\dagger}_{\mathbf{k},l+1}\hat{c}_{\mathbf{k},l}+\Gamma^{\dagger}(\mathbf{k})\hat{c}^{\dagger}_{\mathbf{k},l}\hat{c}_{\mathbf{k},l+1}\right),
    \end{aligned}
\end{equation}
where $h(\mathbf{k})$ represent the monolayer Hamiltonian, $V_l$ the potential energy induced by the displacement field in the $l$-th layer, and $\Gamma(\mathbf{k})$ the hopping matrix between nearest-neighbor layers. To simplify the formulae, we define:
\begin{equation}\label{JJ}
    \begin{aligned}[b]
        \hat{J}_{\pm}(\mathbf{k}) = \sum_{l=1}^{N-1} \left( \Gamma(\mathbf{k}) \hat{c}^{\dagger}_{\mathbf{k},l+1} \hat{c}_{\mathbf{k},l} \pm \Gamma^{\dagger}(\mathbf{k}) \hat{c}^{\dagger}_{\mathbf{k},l} \hat{c}_{\mathbf{k},l+1} \right).
    \end{aligned}
\end{equation}
These definitions will be used in the following sections. Alternatively, we can rewrite $X_{ab}=-\frac{1}{\hbar^2}  \left\langle[\hat{r}_a,[\hat{r}_b,\hat{H}(\mathbf{k})]]\right\rangle_F$. Using the Jacobi identity $[\hat A,[\hat B,\hat C]]+[\hat B,[\hat C,\hat A]]+[\hat C,[\hat A,\hat B]]=0$, we can immediately verify that $X_{ab}=X_{ba}$. 

For $(a,b)=(x,y)$, the in-plane  tensor can be calculated by splitting the position operator into its intra-band and inter-band components, denoted by $\hat \Rv$ and $\hat \Xv$ respectively: $\hat \rv = \hat \Rv+\hat \Xv$. Using the well-known representations
$[\Rv]_{n\kv,n'\kv'}=(i\partial_\kv \delta_{\kv,\kv'}+i\langle u_{n\kv}|\partial_\kv u_{n\kv}\rangle \delta_{\kv,\kv'})\delta_{n,n'}$ and $[\Xv]_{n\kv,n'\kv'}=i\langle u_{n\kv}|\partial_\kv u_{n'\kv}\rangle (1-\delta_{n,n'})$ we arrive at~\cite{supp}
\begin{equation}\label{Drude}
    \begin{aligned}[b]
    X_{ab}=\frac{1}{\hbar}\sum_{n\mathbf{k}}f_{n\mathbf{k}}\frac{\partial v^b_{nn}(\mathbf{k})}{\partial k_a}=\frac{1}{\hbar^2}\sum_{n\mathbf{k}}f_{n\mathbf{k}}\frac{\partial^2 \epsilon_{n}(\mathbf{k})}{\partial k_a \partial k_b}.
    \end{aligned}
\end{equation}
Thus, $X_{ab}$ is the Fermi volume integral of the effective inverse mass tensor $\left[\frac{1}{m^*}\right]_{ab}=\frac{1}{\hbar^2}\frac{\partial^2 \epsilon_{n}(\mathbf{k})}{\partial k_a \partial k_b}$. This is closely related to Drude weight $D_{ab}$, as expressed by the relationship $D_{ab} = 2\pi e^2 X_{ab}$ in metal band theory~\cite{Resta_2018}. This relationship underscores the fundamental connection between the two quantities that characterize the electrical properties of a metal: the Drude weight and the effective mass. 

In the non-periodic directions of the layered system, the position operator is defined as
\begin{equation}\label{}
    \begin{aligned}[b]
        \hat{r}_z=\sum_{l,\mathbf{k}}z_l\hat{c}^{\dagger}_{\mathbf{k},l}\hat{c}_{\mathbf{k},l}.
    \end{aligned}
\end{equation}
where $z_l$ denotes the position of $l$-th layer on the $z$-axis. To simplify the expression, we assume that the spacing between layers is  $z_{l+1}-z_{l}=d$ for all $l$. Thus, $X_{zz}$ is given by
\begin{equation}\label{X-Diagonal}
    \begin{aligned}[b]
        X_{zz}=-\frac{d^2}{\hbar^2}\sum_{n\mathbf{k}}f_{n\mathbf{k}}\mel{u_{n\mathbf{k}}}{\hat{J}_{+}(\mathbf{k})}{u_{n\mathbf{k}}},
    \end{aligned}
\end{equation}
and the off-diagonal elements associated with the out-of-plane direction are given by
\begin{equation}\label{X-OffDiagonal}
    \begin{aligned}[b]
        X_{za}=-\frac{d}{\hbar^2}\sum_{n\mathbf{k}}f_{n\mathbf{k}}\mel{u_{n\mathbf{k}}}{\frac{\partial \hat{J}_{-}(\mathbf{k})}{\partial k_a}}{u_{n\mathbf{k}}}.
    \end{aligned}
\end{equation}
From Eq.~(\ref{JJ}) we see that the magnitudes of $X_{zz}$ and $X_{za}$  are controlled by the interlayer coupling $\Gamma(\mathbf{k})$. The details of the derivation are given in~\cite{supp}. When the layered system has sufficiently high symmetry in the plane~\cite{supp}, the off-diagonal components $X_{za}$ vanish.

\textit{Model system with layer stacking.--}
For a simple illustration of the theory, we consider the following $4\times4$ tight-binding model of bilayer graphene (BLG)~\cite{McCann2013}:
\begin{equation}
    \begin{aligned}[b]
        H_{0}(\mathbf{k})=-\gamma_0\mathbf{h}(\mathbf{k})\cdot\bm{\sigma}\tau_0+\frac{\gamma_1}{2}\left(\sigma_x\tau_x+\sigma_y\tau_y\right)+V\sigma_0\tau_z,
    \end{aligned}
\end{equation}
where $\gamma_0$ is the intra-layer nearest neighbor hopping, $\gamma_1$ is the interlayer hop[ping, $V$ is the inter-layer bias, $\mathbf{h}(\mathbf{k})=(F(\mathbf{k}),G(\mathbf{k}))$ is the 2D vector with $F(\mathbf{k})=\Re f(\mathbf{k})=\sum_{n}\cos(\bm{\delta}_n\cdot\mathbf{k})$ and $G(\mathbf{k})=-\Im f(\mathbf{k})=-\sum_{n}\sin(\bm{\delta}_n\cdot\mathbf{k})$. 
Here, $\bm{\sigma}$ is the sublattice pseudo-spin variable and $\bm{\tau}$ acts on the layer index:
$\tau_z = d[\hat{c}^\dagger_1(\kv)\hat{c}_1(\kv)-\hat{c}^\dagger_2(\kv)\hat{c}_2(\kv)]/2$,
$\tau_+ = \hat{c}^\dagger_1(\kv) \hat{c}_2(\kv)$ and $\tau_- = \hat{c}^\dagger_2(\kv) \hat{c}_1(\kv)$. 

Using Eqs.~(\ref{JJ}) and (\ref{X-OffDiagonal}), we can directly calculate the $X_{ab}$ tensor. Specifically, with $\hat r_z = \frac{d}{2}\sigma_0\tau_z$ in this bilayer system, we have
\begin{equation}
    \begin{aligned}[b]   
    X_{zz}=-\frac{\gamma_1 d^2}{2\hbar^2}\langle\left(\sigma_x\tau_x+\sigma_y\tau_y\right)\rangle_F.
    \end{aligned}
\end{equation}
which is proportional to the interlayer hopping $\gamma_1$. 

\begin{figure}[t]
    \includegraphics[width=0.48\textwidth]{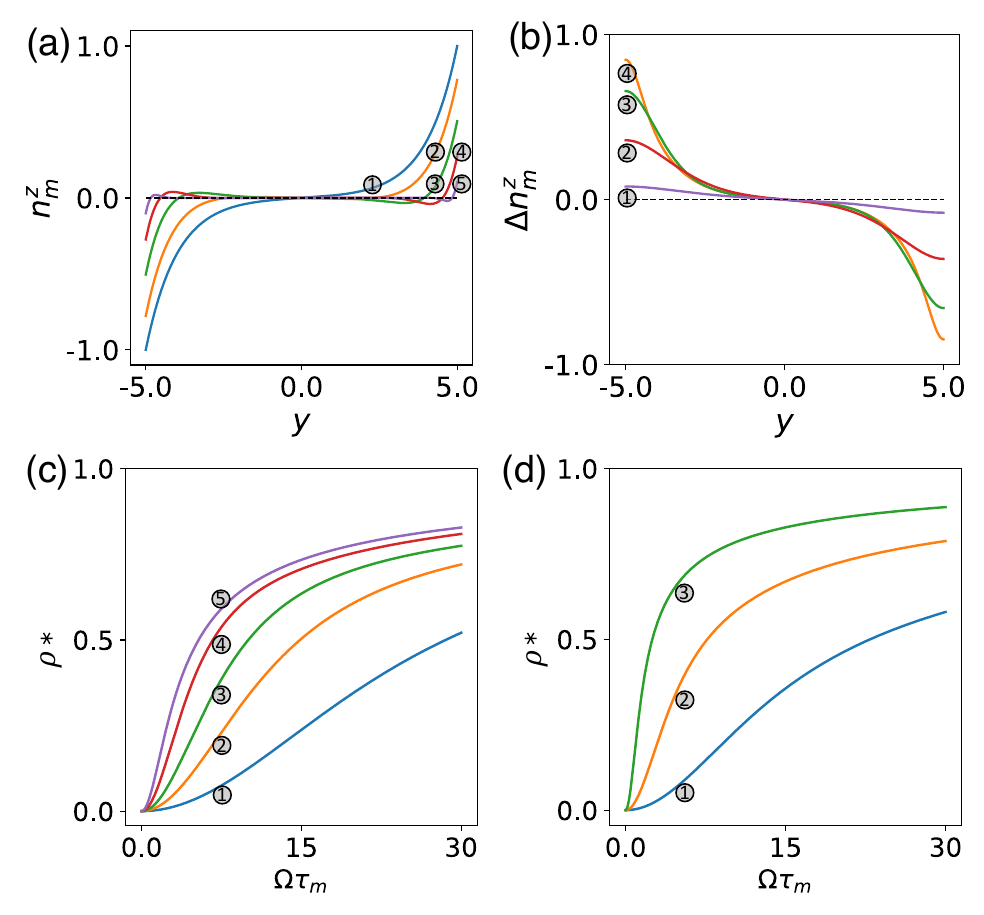}
    \caption{\label{fig2} (a) Spatial dependence of the OMM density  from Eq.~\eqref{nz_dis} for different $B$-field values: $\circledtext{1}$ $B=0$, $\circledtext{2}$ $B=1.0$, $\circledtext{3}$ $B=2.5$, $\circledtext{4}$ $B=7.5$, $\circledtext{5}$ $B=50$. (b) Dependence of $\Delta n^z_{\mathbf{m}}(y)=n^z_{\mathbf{m}}(\theta,y)-n^z_{\mathbf{m}}(0,y)$ on the anisotropy parameter $\theta$ for a fixed $\Omega\tau_m=2$. $\circledtext{1}$ $\theta=10^{-2}$, $\circledtext{2}$ $\theta=10^{-1}$, $\circledtext{3}$ $\theta=1$, $\circledtext{4}$ $\theta=20$. (c) Normalized relative HMR vs $\Omega\tau_m$  for different values of the sample width $L^*=L/L_m$. $\circledtext{1}$ $L^*=1.0$, $\circledtext{2}$ $L^*=1.5$, $\circledtext{3}$ $L^*=2.0$, $\circledtext{4}$ $L^*=3.0$, $\circledtext{5}$ $L^*=10$. (d) Normalized relative HMR vs $\Omega\tau_m$  for different values of the anisotropy parameter $\theta$.  $\circledtext{1}$ $\theta=10^{-2}$, $\circledtext{2}$ $\theta=10^{-1}$, $\circledtext{3}$ $\theta=1$.}
\end{figure}

In Fig.~\ref{fig3} (a), we plot the eigenvalues of the inverse mass tensor $X$ versus the Fermi level, vis-\`a-vis, the band structure of the model (green curves). Due to the $\hat{c}_3$ in-plane symmetry, $X_{xx}$ and $X_{yy}$ are identical and vanish when the Fermi energy is in the insulating gap. However,  $X_{zz}$  is significantly smaller (for realistic values of the interlayer hopping), reflecting a much larger effective mass for interlayer motion. The impact of this reduced mass on magnetoresistance is clearly visible in Fig.~\ref{fig2} (d). 

\textit{Orbital Hanle magnetoresistance --} We closely parallel the theory of Hanle magnetoresistance laid out by Dyakonov for spintronics.\cite{Dyakonov1994}  The OHE creates an accumulation of orbital magnetic moment near the system edges. This accumulation contributes to an electric current that flows parallel to the edges, causing a slight decrease in resistance. The accumulation of orbital moment and the resulting change in resistance can be modulated by an in-plane magnetic field. Following Dyakonov~\cite{Dyakonov1994, Wang2023}, we start with a drift-diffusion equation for the electric current:
\begin{equation}\label{d-d2}
    \begin{aligned}[b]
        \mathbf{J}=\mathbf{J}^{(0)}+\mu_e\alpha\mathbf{E}\wedge \mathbf{n}_m+D\alpha\mathbf{\nabla}\times\mathbf{n}_m\,,
    \end{aligned}
\end{equation}
where $D$ is the diffusion constant, $\mu$ the drift mobility, and $\alpha$ the orbital Hall angle.~\footnote{The $\wedge$ symbol denotes an external product between spatial and spin indices.}
The last term on the right-hand side says that a charge current arises from the curl of the OMM density: this current has a component parallel to the edge and hence contributes to the sample resistance if the out-of-plane component of the OMM varies as a function of the distance from the sample edge. Up to this point, our formulation coincides with that of Ref.~\cite{Dyakonov1994}. See~\cite{supp} for the derivation of Eq.~(\ref{d-d2}).   

Solving Eq.~\eqref{EOM_Mac} in steady state ($\frac{\partial \mathbf{n}_{\mathbf{m}}}{\partial t}=0$), we obtain the spatial distributions of the OMM density. We set the electric field and the magnetic field along the $x$ axis: $\mathbf{E}=E \mathbf{e}_x$ and $\mathbf{B}=B \mathbf{e}_x$. The system is finite along the $y$-direction with edges at $y=\pm \frac{L}{2}$. We assume, for simplicity, that the system has sufficiently high symmetry (for example, $\hat{C}_3$ rotational symmetry) to guarantee the vanishing of all the off-diagonal components of the inverse mass tensor~\cite{supp}.  With these assumptions, we arrive at the solution that satisfies the boundary conditions $J^y_{m_z}=0$ and $J^y_{m_y}=0$: 
\begin{equation}\label{nz_dis}
    \begin{aligned}[b]
    n^z_m(y) = -\frac{e \sigma_{OH} E}{D}\Re{\frac{\sinh{\lambda y}}{\lambda\cosh{\frac{\lambda L}{2}}}}.
    \end{aligned}
\end{equation}
where $\lambda = \frac{\sqrt{1 + i \sqrt{\theta \phi^2}}}{L_m}$ with $L_m=\sqrt{D\tau_m}$, $\phi=\Omega\tau_m$, $\theta=\frac{X_{zz}}{X_{yy}}$ and $\Omega=eBX_{yy}$. Here $L_m$ is the OMM diffusion length and $\sigma_{OH}=\alpha \mu_e n$ is orbital Hall conductivity. The parameter $\phi$ is a measure of the strength of the magnetic field and $\theta$ is a measure of anisotropy, i.e. $\theta=X_{zz}/X_{yy}$, with $\theta=1$ describing the isotropic case (equivalent to the spin case) and $\theta=0$ corresponding to the case of zero interlayer coupling, that is, OMM locked to the $z$ axis.  The solution is plotted in Fig.~\ref{fig2} (a),  where we use $\alpha = 0.016$ and $\tau_m=2$ ps and $L_m= 2$ nm, as suggested in Ref.~\cite{Sala2023}). We notice that these parameters could be larger in 2D Dirac materials due to higher electron mobilities~\cite{Dean2010}.

Using Eq.~\eqref{d-d2}, we calculate the total current flowing in the $x$ direction as
\begin{equation}
 I_{tot}=I_0+ \Delta I\,,  
\end{equation}
where $I_0=-e\mu_e n E L$ is the current without OMM-induced corrections and
\begin{equation}\label{}
    \begin{aligned}[b]
        \Delta I=\int^{\frac{L}{2}}_{-\frac{L}{2}} dy \Delta J_x(y)=\alpha^2 I_0\Re{\frac{2}{\lambda L}\tanh{\frac{\lambda L}{2}}}
        \end{aligned}
\end{equation}
is the contribution of the OMM accumulations. The fractional change in resistance, defined as $\rho(B) \equiv \frac{R(B) - R_0}{R(B)} = -\frac{\Delta I}{I_0}$, depends on the magnetic field $B$, and has explicit form
\begin{equation}\label{RhoB}
    \begin{aligned}[b]
        \rho(B)=-\alpha^2 \Re{\frac{2}{\lambda L}\tanh{\frac{\lambda L}{2}}}.
        \end{aligned}
\end{equation}
Here, $R(B)$ represents the resistance including the effect of the OMM accumulation under the influence of a magnetic field, while $R_0$ is the intrinsic resistance of the material without the OMM accumulation effect. For $B \to \infty$, the resistance induced by the accumulation of OMM is completely suppressed, and thus $R(\infty) = R_0$ and $\rho(\infty)=0$.

It is convenient to define a normalized relative HMR, $\rho^*(B)$, as follows  
\begin{equation}\label{}
    \begin{aligned}[b]
        \rho^{*}(B)=\frac{\rho(B)-\rho(0)}{\rho(\infty)-\rho(0)}.
    \end{aligned}
\end{equation}
This quantity indicates the relative change in resistance between its zero-field value and its large field value ($B \to \infty$). Figures~\ref{fig2} (c) and (d) show the plot of $\rho^*$ as a function of the magnetic field strength $\Omega\tau_m$ for different parameters. In Fig.~\ref{fig2} (c), we show the width dependence of HMR using the parameter $L^*=\frac{L}{L_m}$. When the diffusion length $L_m$ is much smaller than the width of the system, $\rho^*$ changes rapidly, indicating a high sensitivity of the MR response. Conversely, when $L_m$ is comparable to or larger than the system width, the change in $\rho^*$ is less pronounced. In this regime, the impact of the magnetic field on resistance is diminished as the system's geometry no longer facilitates the enhanced scattering effects that occur when the width is much larger than the diffusion length. As a result, the HMR response is less sensitive to the magnetic field.

The  dependence of $\rho^*$ on anisotropy is illustrated in Fig.~\ref{fig2}(d). We present the magnetoresistance for $\theta = 10^{-2}$, $10^{-1}$, and $1$ respectively. As $\theta$ decreases, the sensitivity of the MR also decreases. This diminished sensitivity with smaller $\theta$ values is due to the larger inertia of the inter-layer motion, which is captured by the averaged effective inverse mass tensor $X_{cd}$. Notably, $\theta = 1$ corresponds to the isotropic case, providing a direct comparison to spin-related phenomena~\cite{Dyakonov1994,Casanova2009}.

\textit{Discussion and outlook --}
Although designed for a completely different class of systems, our phenomenological theory aligns well with the experimental results reported in Ref.~\cite{Sala2023} for  isotropic polycrystalline three-dimensional Mn films, where the role of the layer width $L$ is played by the film thickness.  The observed dependence of the HMR on thickness (plotted in Fig. 4a of Ref.~\cite{Sala2023})  demonstrates a size effect that is consistent with the width dependence of $[\rho(B)-\rho(0)]$ calculated from Eq.~(\ref{RhoB}).  
Specifically, when the width is much smaller than the diffusion length ($L \ll L_m$), OMM depolarization is dominated by diffusion processes, but when the width is much larger than the diffusion length ($L \gg L_m$), direct OMM relaxation processes become the primary mechanism of depolarization~\cite{Dyakonov1994}. The crossover between the two regimes produces a maximum in HMR as a function of film thickness.

The real test of the theory, however, will come from its application to layered structure with highly anisotropic properties. In spite of the large anisotropy, which is clearly detrimental for HMR, We have found that the response $[\rho(B) - \rho(0)]$ in bilayer graphene (BLG) is an order of magnitude higher than the value reported in the three-dimensional Mn film~\cite{Sala2023}. We attribute this enhancement to the small effective mass of electrons in BLG  $m^* = 0.041 m_e$, which is 675 times smaller than that in manganese)~\cite{ashcroft1976, Zou2011} and the correspondingly higher mobility. Consequently, the precession frequency of orbital moments around the out-of-plane axis in BLG is 675 times higher than $\Omega_{Mn}$ in manganese. Taking into account the effect of the anisotropy (with $\theta =\frac{X_{zz}}{X_{yy}} \simeq 10^{-2}$), we find that the effective precession frequency $\Omega^*_{BLG} = \sqrt{\theta} \Omega_{BLG}$ is ``only'' 67.5 times $\Omega_{Mn}$. In spite of this, our layered 2D system is still expected to exhibit a substantially larger magneto-resistance than Mn. Recently, it has been proposed that a large OMM could exist in twisted bilayer graphene, potentially providing an excellent platform for realizing OMM dynamics in stacked 2D materials~\cite{Li2020,He2020}. 

Looking ahead, we predict that a vertical displacement field applied by gating can be used to tune the in-plane components of the Drude weight and hence the orbital HMR of layered systems. Fig.~\ref{fig3}(b) shows  plots of the anisotropy ratio $\theta$ versus $V$. The sensitivity of $\theta$ to the value of $V$ suggests that the orbital HMR may be more controllable than traditional spin HMR and offer  new possibilities for advanced magnetic sensors and memory devices. More generally, we believe that our theory will have wide applicability and provide a robust and versatile tool for the interpretation of experiments dealing with the orbital magnetic moment dynamics of layered systems.

\begin{acknowledgements}
    We acknowledge J. C. W. Song for useful comments on the manuscript. This research is supported by the Ministry of Education, Singapore, under its Research Centre of Excellence award to the Institute for Functional Intelligent Materials (I-FIM, project No. EDUNC-33-18-279-V12).
\end{acknowledgements}

\bibliographystyle{apsrev4-2}
\bibliography{orb_Hall.bib}

\onecolumngrid
\clearpage
\begin{center}
\textbf{\large Supplemental Materials}
\end{center}

\setcounter{equation}{0}
\setcounter{figure}{0}
\setcounter{table}{0}
\setcounter{page}{1}
\makeatletter
\renewcommand{\theequation}{S\arabic{equation}}
\renewcommand{\thefigure}{S\arabic{figure}}
\renewcommand{\bibnumfmt}[1]{[S#1]}
\renewcommand{\citenumfont}[1]{#1}

\section{Phenomenological Drift-Diffusion Theory}
This section is for the derivation of Eq.~\eqref{d-d2} in the main text. Following Dyakonov~\cite{Dyakonov1994, Wang2023}, we start with phenomenological expressions for the 
in-plane ``zero-th order'' charge current $J^{(0)}_i$ and the OMM current $[J_m]_{i}^{a,(0)}$ (the flow of the $a$ component of the orbital moment in the $i$ spatial direction) in the presence of an electric field $\mathbf{E}$
\begin{equation}\label{d-d}
    \begin{aligned}[b] 
     J^{(0)}_{i}&=e\mu_e E_i n+eD\nabla_i n,\\
        [J_m]_i^{a,(0)}&=-\mu_e E_i n^a_m-D\nabla_i n^a_m,\\
    \end{aligned}
\end{equation}
where $i,j,k...$ are two-dimensional cartesian indices for spatial directions, $a,b,c...$ are three-dimensional cartesian indices for OMM directions, $n$ is the electron density, $n^a_m$ is the $a$  component of the OMM density, $\mu_e$ is the electronic mobility and $D$ is the  diffusion constant. The above equations ignore the existence of the orbital Hall effect, hence the $(0)$ superscript.  

When OHE and its inverse are included the charge current becomes an additional source of OMM current and vice-versa according to the equations 
\begin{equation}\label{d-d-2}
    \begin{aligned}[b]         
        J^{(1)}_i&=-\alpha\epsilon_{ijc}[J_m]_{j}^{c,(0)},\\
        [J_m]_{i}^{a,(1)}&=\alpha\epsilon_{iak}J^{(0)}_{k},
    \end{aligned}
\end{equation}
where $\alpha$ is the orbital Hall angle which describes the strength of the OHE (sums over repeated indices are implied).
Importantly, this phenomenological description of the OHE ensures that any ``undergap'' component of the transverse OMM current is automatically excluded from consideration because it does not arise from the conversion of a longitudinal charge current~\cite{Kazantsev2024}. Thus the present theory is relevant only for metallic states.
Combining Eqs.~(\ref{d-d}) and ~(\ref{d-d-2}) up to the first order terms, we obtain the total longitudinal current
\begin{equation}\label{}
    \begin{aligned}[b]         
        J_i=&J^{(0)}_i+J^{(1)}_i\\
        =&J^{(0)}_i+\alpha \mu_e\epsilon_{ijc}E_j n^c_m+\alpha D\epsilon_{ijc}\nabla_j n^c_m,
    \end{aligned}
\end{equation}
Thus, the vector expression form is given by
\begin{equation}\label{}
    \begin{aligned}[b]         
        \mathbf{J}=\mathbf{J}^{(0)}+\alpha\mu_e\mathbf{E}\wedge \mathbf{n}_m+\alpha D\nabla \times \mathbf{n}_m\,,
    \end{aligned}
\end{equation}
i.e., Eq.~\eqref{d-d2} of the main text.

\section{Dynamics of the orbital magnetization}\label{s7}
In this section we derive Eqs.~(\ref{EOM_Mac}) and (\ref {X_tensor}) of the main text. We start with a general one-particle Hamiltonian with the external fields
\begin{equation}
    \begin{aligned}[b]
        \hat{H}_{em}=\hat H_0+\sum_p\left[-e\mathbf{E}\cdot\hat{\mathbf{r}}_p+\mathbf{B}\cdot \hat{\mathbf{m}}_p\right],
    \end{aligned}
\end{equation}
where $\mathbf{E}$ is the electric field, and $\mathbf{B}$ is the magnetic field. $\hat{H}_0$ is the zero-field Hamiltonian. The index $p$ runs over the electrons.  The orbital magnetic moment operator (for one electron)  has the form
\begin{equation}
    \begin{aligned}[b]
        \hat{m}_a=\frac{-e}{4}\epsilon^{abc}\{\hat{r}_b,\hat{v}_c\}=\frac{-e}{2}\epsilon^{abc}\hat{r}_b\star\hat{v}_c.
    \end{aligned}
\end{equation}
The velocity operator is given by $\hat{\mathbf{v}}=\frac{1}{i\hbar}[\hat{\mathbf{r}},\hat{H}_{em}]$. Here, we use the Einstein summation convention, where repeated indices imply summation over those indices. The commutator between different components of the OMM operator can be written as
\begin{equation}\label{}
    \begin{aligned}[b][\hat{m}_a,\hat{m}_b]=\frac{e^2}{4}\epsilon^{akl}\epsilon^{bk'l'}\hat{r}_{k}\star\hat{r}_{k'}\star[\hat{v}_l,\hat{v}_{l'}]+\frac{e^2}{4}\left(\epsilon^{akl}\epsilon^{bk'l'}-\epsilon^{ak'l'}\epsilon^{bkl}\right)\hat{r}_{k'}\star[\hat{r}_k,\hat{v}_{l'}]\star\hat{v}_{l}.
    \end{aligned}
\end{equation}
The commutator of two components of the velocity vanishes at a zero magnetic field, and thus it can be neglected in the derivation of the equations of motion to linear order in $B$. Using the definition $[\hat{r}_k,\hat{v}_{l'}]=i\hbar X_{kl'}$, the second term on the right-hand side can be simplified to
\begin{equation}\label{}
    \begin{aligned}[b]
        &\frac{i\hbar e^2}{4}\left(\epsilon^{akl}\epsilon^{bk'l'}-\epsilon^{ak'l'}\epsilon^{bkl}\right)\hat{r}_{k'}\star\hat{v}_{l} X_{kl'}\\
        &=\frac{i\hbar e^2}{4}\epsilon^{abc}\left(\epsilon^{kk'c}\hat{r}_{k'}\star\hat{v}_lX_{kl}+\epsilon^{k'lc}\hat{r}_{k'}\star\hat{v}_lX_{kk}+\epsilon^{l'kc}\hat{r}_{l}\star\hat{v}_lX_{kl'}+\epsilon^{ll'c}\hat{r}_{k}\star\hat{v}_lX_{kl'}\right)\\
        &=\frac{-i\hbar e}{2}\epsilon^{abk}X_{kl}\hat{m}_l.
    \end{aligned}
\end{equation}
In the derivation, we have employed the identity of the Levi-Civita symbol
\begin{equation}\label{}
    \begin{aligned}[b]
    \epsilon^{akl}\epsilon^{bk'l'}=\delta_{ab}\left(\delta_{kk'}\delta_{ll'}-\delta_{kl'}\delta_{lk'}\right)-\delta_{ak'}\left(\delta_{kb}\delta_{ll'}-\delta_{kl'}\delta_{bl}\right)+\delta_{al'}\left(\delta_{kb}\delta_{k'l}-\delta_{kk'}\delta_{bl}\right).
    \end{aligned}
\end{equation}
Thus, we arrive at 
\begin{equation}\label{}
    \begin{aligned}[b][\hat{m}_a,\hat{m}_b]=\frac{e^2}{4}\epsilon^{akl}\epsilon^{bk'l'}\hat{r}_{k}\star\hat{r}_{k'}\star[\hat{v}_l,\hat{v}_{l'}]-\frac{i\hbar e}{2}\epsilon^{abk}X_{kl}\hat{m}_l,
    \end{aligned}
\end{equation}
and
\begin{equation}\label{}
    \begin{aligned}[b][\hat{m}_a,\hat{m}_bB_b]&=\frac{e^2}{4}\epsilon^{akl}\epsilon^{bk'l'}\hat{r}_{k}\star\hat{r}_{k'}\star[\hat{v}_l,\hat{v}_{l'}]B_j-\frac{i\hbar e}{2}\epsilon^{abk}B_bX_{kl}\hat{m}_l\\
    &\approx-\frac{i\hbar e}{2}\epsilon^{abk}B_bX_{kl}\hat{m}_l.
    \end{aligned}
\end{equation}
Here again we retain only terms that contribute to the equation of motion in linear order in $B$.
Now the equation of motion for the OMM density operator is
\begin{equation}\label{EOM}
    \begin{aligned}[b]
        i\hbar\frac{\partial }{\partial t}\hat{n}^a_{m}(\mathbf{r})=&\left[\hat{n}^a_{m}(\mathbf{r}),\hat{H}_E\right]+\left[\hat{n}^a_{m}(\mathbf{r}),\hat{\mathbf{m}}\cdot\mathbf{B}\right]\\
        =&\left[\hat{n}^a_{m}(\mathbf{r}),\hat{H}_E\right]+\sum_p\hat{m}^a_{p}
        \star\left[\delta(\mathbf{r}-\hat{\mathbf{r}}_p),\hat{\mathbf{m}}_p\cdot\mathbf{B}\right]+\sum_p\left[\hat{m}^a_{p},\hat{\mathbf{m}}_p\cdot\mathbf{B}\right]\star\delta(\mathbf{r}-\hat{\mathbf{r}}_p).
    \end{aligned}
\end{equation}
 The first and third terms can be cast as the spatial divergence of the OMM current density and the orbital torque density, respectively. We can rewrite $\frac{1}{i\hbar}\left[\hat{n}^a_{m}(\mathbf{r}),\hat{H}_E\right]+\frac{1}{i\hbar}\sum_p\hat{m}^a_{p}
        \star\left[\delta(\mathbf{r}-\hat{\mathbf{r}}_p),\hat{\mathbf{m}}_p\cdot\mathbf{B}\right]\approx-\nabla\cdot\hat{\mathbf{J}}_{m}^a-\frac{\hat{n}^a_{m}(\mathbf{r})}{\tau_m}$.
where the orbital current density operator is given by 
\begin{equation}\label{}
    \begin{aligned}[b]
    \hat{\mathbf{J}}_{m}^a=\sum_{p}\hat{m}^{a}_p\star\delta\left(\mathbf{r}-\hat{\mathbf{r}}_p\right)*\hat{\mathbf{v}}_p\,.
    \end{aligned}
\end{equation}

Thus, the full equation of motion for the OMM density operator is 
\begin{equation}\label{s_EOM}
    \begin{aligned}[b]
         \frac{\partial }{\partial t}\hat{n}^a_m=&-\nabla \cdot \hat{\mathbf{J}}_m^a-\frac{\hat{n}^a_m}{\tau_m}-\frac{e}{2}\epsilon^{abc}B_b X_{cd} \hat{n}_m^d\,,
    \end{aligned}
\end{equation}
which is Eq.~(\ref {X_tensor}) in the text. 

The equation of motion for the OMM density (orbital magnetization) can be directly obtained by averaging the operator equation over the non equilibrium state 
\begin{equation}\label{s_EOM_Mac}
    \begin{aligned}[b]
        \frac{\partial }{\partial t}n^a_m=&-\nabla \cdot \mathbf{J}_m^a-\frac{n^a_m}{\tau_m}-\frac{e}{2}\epsilon^{abc}B_b X_{cd} n_m^d\,,
    \end{aligned}
\end{equation}
which is Eq.~(\ref{EOM_Mac}) in the text.

\section{Obtaining the OMM density in the steady state}
Here we derive Eq.~\eqref{nz_dis} of the main text. 
The coupled drift-diffusion equations for the OMM density in the steady state are
\begin{equation}\label{}
    \begin{aligned}[b]
        -D\frac{\partial^2}{\partial y^2}n^z_{\mathbf{m}}-\frac{n^z_{\mathbf{m}}}{\tau_m}-\frac{e}{2}B_x X_{yy}n^y_{\mathbf{m}}=0,\\
        -D\frac{\partial^2}{\partial y^2}n^y_{\mathbf{m}}-\frac{n^y_{\mathbf{m}}}{\tau_m}+\frac{e}{2}B_x X_{zz}n^z_{\mathbf{m}}=0.
    \end{aligned}
\end{equation}
By inserting the first equation into the second one, one can eliminate $n^y_{\mathbf{m}}$ and obtain the differential equation $n^z_{\mathbf{m}}$:
\begin{equation}
    \begin{aligned}[b]
    \frac{D^2 \tau_m^2}{\Omega^2 \tau_m^2} \frac{\partial^4 n^z_m}{\partial y^4} - \frac{2D \tau_m}{\Omega^2 \tau_m^2} \frac{\partial^2 n^z_m}{\partial y^2} + \left( \frac{\Omega_z}{\Omega} + \frac{1}{\Omega^2 \tau_m^2} \right) n^z_m = 0
    \end{aligned}
\end{equation}
where $\Omega=eBX_{yy}$, and $\Omega_z=eBX_{zz}$. We set $\frac{\Omega_z}{\Omega}=\theta$, $D\tau_m=L^2_m$ and $\Omega\tau_m=\phi$. The $L_m$ is the OMM diffusion length. This simplifies to
\begin{equation}
    \begin{aligned}[b]
    \frac{L_m^4}{\phi^2} \frac{\partial^4 n^z_m}{\partial y^4} - \frac{2 L^2_m}{\phi^2} \frac{\partial^2 n^z_m}{\partial y^2} + \left( \theta + \frac{1}{\phi^2} \right) n^z_m = 0
    \end{aligned}
\end{equation}
The general solution can be obtained using the characteristic equation. Let's start by assuming a solution of the form $n^z_m(y) = e^{\lambda y}$.
Substituting it into the original differential equation gives us the characteristic equation:
\begin{equation}
    \begin{aligned}[b]
    \frac{L_m^4}{\phi^2} \lambda^4 - \frac{2 L_m^2}{\phi^2} \lambda^2 + \left( \theta + \frac{1}{\phi^2} \right) = 0.
    \end{aligned}
\end{equation}
Therefore, we obtain the following
\begin{equation}
    \begin{aligned}[b]
    \lambda^2 = \frac{1 \pm i \sqrt{\theta \phi^2}}{L_m^2}.
    \end{aligned}
\end{equation}
Taking the square root, we get the following
\begin{equation}
    \begin{aligned}[b]
    \lambda_1 = \pm \sqrt{\frac{1 + i \sqrt{\theta \phi^2}}{L_m^2}}, \quad \lambda_2 = \pm \sqrt{\frac{1 - i \sqrt{\theta \phi^2}}{L_m^2}}.
    \end{aligned}
\end{equation}
Thus, the general solution to the differential equation is:
\begin{equation}
    \begin{aligned}[b]
    n^z_m(y) =& \hat{c}_1 e^{\frac{\sqrt{1 + i \sqrt{\theta}\abs{ \phi}}}{L_m} y} + \hat{c}_2 e^{-\frac{\sqrt{1 + i \sqrt{\theta}\abs{ \phi}}}{L_m} y} + \hat{c}_3 e^{\frac{\sqrt{1 - i \sqrt{\theta}\abs{ \phi}}}{L_m} y} + \hat{c}_4 e^{-\frac{\sqrt{1 - i \sqrt{\theta}\abs{ \phi}}}{L_m} y}\\
    =&(\hat{c}_1-\hat{c}_2)\sinh{\lambda_1 y}+(\hat{c}_1+\hat{c}_2)\cosh{\lambda_1 y}+(\hat{c}_3-\hat{c}_4)\sinh{\lambda_2 y}+(\hat{c}_3+\hat{c}_4)\cosh{\lambda_2 y}.
    \end{aligned}
\end{equation}
where we choose $\lambda_1 =\sqrt{\frac{1 + i \sqrt{\theta \phi^2}}{L_m^2}}$ and $\lambda_2 =\sqrt{\frac{1 - i \sqrt{\theta \phi^2}}{L_m^2}}$. Here, \(\hat{c}_1\), \(\hat{c}_2\), \(\hat{c}_3\), and \(\hat{c}_4\) are arbitrary constants determined by boundary conditions. 

We can simplify the function further
\begin{equation}
    \begin{aligned}[b]
    n^z_m(y) =\hat{c}^1_a\Re{\sinh{\lambda_1 y}}+i\hat{c}^2_a\Im{\sinh{\lambda_1 y}}+\hat{c}^3_s\Re{\cosh{\lambda_1 y}}+i\hat{c}^4_s\Im{\cosh{\lambda_1 y}}\,,
    \end{aligned}
\end{equation}
where we have used the fact $\Re{\sinh{\lambda_1 y}}=\Re{\sinh{\lambda_2 y}}$, $\Im{\sinh{\lambda_1 y}}=-\Im{\sinh{\lambda_2 y}}$, $\Re{\cosh{\lambda_1 y}}=\Re{\cosh{\lambda_2 y}}$, and $\Im{\cosh{\lambda_1 y}}=-\Im{\cosh{\lambda_2 y}}$. $\hat{c}_a$ and $\hat{c}_s$ are combinations of \(\hat{c}_1\), \(\hat{c}_2\), \(\hat{c}_3\), and \(\hat{c}_4\). Given the boundary conditions $J^{yz}=0$ and $J^{yy}=0$, we have
\begin{equation}
    \begin{aligned}[b]
    -D\frac{\partial n^z_m}{\partial y}\vert_{y=\left(\frac{-L}{2},\frac{L}{2}\right)}=&e\alpha_{OH}\mu_e n E,\\
    D\frac{\partial n^y_m}{\partial y}\vert_{y=\left(\frac{-L}{2},\frac{L}{2}\right)}=&0.
    \end{aligned}
\end{equation}
If we assume the solution \(n^z_m(y)\) to be antisymmetric, that is, \(n^z_m(-y) = -n^z_m(y)\), then we have \(\hat{c}^3_s = \hat{c}^4_s=0\). To simplify the solution, we set $\hat{c}^1_a=\hat{c}^2_a$ and $n^z_m(y)=\hat{c}_a\sinh{\lambda_1 y}$. Using the boundary conditions, we obtain
\begin{equation}
    \begin{aligned}[b]
    \hat{c}_a=-\frac{e \alpha_{OH} \mu_e n E}{D\lambda_1\cosh{\frac{\lambda_1L}{2}}}.
    \end{aligned}
\end{equation}
Thus, the antisymmetric solution is
\begin{equation}
    \begin{aligned}[b]
    n^z_m(y) = -e \alpha_{OH} \mu_e n E\frac{\sinh{\lambda_1 y}}{D\lambda_1\cosh{\frac{\lambda_1L}{2}}}.
    \end{aligned}
\end{equation}
This solution satisfies both the differential equation and the boundary conditions for the antisymmetric case.

\section{Calculation of the inverse mass tensor in layered systems}
Here and in the next section we derive Eqs.~(\ref{Drude}),(\ref{X-Diagonal}), and~(\ref{X-OffDiagonal}) of the main text. 

We start from the tight-binding Hamiltonian of the layered system (with $N$ layers), which is
\begin{equation}\label{}
    \begin{aligned}[b]
    \hat{H} =& \sum_{l=1}^N \sum_{\iv \jv}\left(t_{\iv\jv}+V_l\delta_{\iv\jv}\right)\hat{c}^{\dagger}_{\iv,l} \hat{c}_{\jv,l}
    +\sum_{l=1}^{N-1}\sum_{\iv\jv}\left(\Gamma_{\iv\jv}\hat{c}^{\dagger}_{\iv,l+1} \hat{c}_{\jv,l}+\Gamma^{*}_{\jv\iv}\hat{c}^{\dagger}_{\jv,l} \hat{c}_{\iv,l+1}\right),
    \end{aligned}
\end{equation}
where $\iv$ and $\jv$ are 2D vectors indexing unit cells in a 2D lattice. $t_{\iv\jv}$ is the intralayer hopping, and $\Gamma_{\iv \jv}$ is interlayer hopping between site $\iv$  $l$-th layer and site $\jv$ in $(l+1)$-th layer.
Both $t_{\iv\jv}$ and $\Gamma_{\iv\jv}$ are matrix elements in an internal sublattice space (sites in the unit cell) and depend only on the difference $\iv-\jv$. Switching to the in-layer plane wave basis $\hat{c}^{\dagger}_{\mathbf{k},l}$ (where $\mathbf{k}$ is a two-dimensional wave vector), we get the Hamiltonian in reciprocal space
\begin{equation}\label{}
    \begin{aligned}[b]
        \hat{H}(\mathbf{k}) = \sum_{l=1}^N \left(h(\mathbf{k})+V_lI\right) \hat{c}^{\dagger}_{\mathbf{k},l}\hat{c}_{\mathbf{k},l}
        +\sum_{l=1}^{N-1}\left(\Gamma(\mathbf{k})\hat{c}^{\dagger}_{\mathbf{k},l+1}\hat{c}_{\mathbf{k},l}+\Gamma^{\dagger}(\mathbf{k})\hat{c}^{\dagger}_{\mathbf{k},l}\hat{c}_{\mathbf{k},l+1}\right)
    \end{aligned}
\end{equation}
where $I$ is the identity matrix in the in-layer space, $h(\mathbf{k})$ is hermitian matrix. The in-plane velocity is given by
\begin{equation}\label{}
    \begin{aligned}[b]
        \hat{v}_x=\frac{1}{\hbar}\frac{\hat{H}(\mathbf{k})}{\partial k_x},\quad \hat{v}_y=\frac{1}{\hbar}\frac{\hat{H}(\mathbf{k})}{\partial k_y}.
    \end{aligned}
\end{equation}
The in-plane components of the Drude tensor are well known~\cite{Resta_2018}:
\begin{equation}\label{OurDrude}
    \begin{aligned}[b]
        X_{ab}(\mathbf{k})=-\frac{1}{\hbar^2}\sum_{n\kv}f_{n\kv}\mel {u_{n\kv}}{[\hat r_a,[\hat r_b,\hat{H}(\mathbf{k})]}{u_{n\kv}}= \frac{1}{\hbar^2}\sum_{n\kv}f_{n\kv}\frac {\partial^2 E_n(\kv)}{\partial k_a \partial k_b}\,.
    \end{aligned}
\end{equation}
We will provide our own derivation of this result in the next section. 

We now focus on  $X_{zz}$. To calculate velocity along the $z$ direction, we define the $z$-component of the position operator 
\begin{equation}\label{}
    \begin{aligned}[b]
        \hat{r}_z=\sum_{l,\mathbf{k}}z_l\hat{c}^{\dagger}_{\mathbf{k},l}\hat{c}_{\mathbf{k},l}.
    \end{aligned}
\end{equation}
where $z_l$ is the position of the $l$-th layer along the $z$-axis.  The $z$-component of the velocity is
\begin{equation}\label{}
    \begin{aligned}[b]
        \hat{v}_z=\frac{1}{i\hbar}[\hat{r}_z,\hat{H}(\mathbf{k})]=\frac{1}{i\hbar}\sum^{N-1}_{l}(z_{l+1}-z_{l})\left(\Gamma(\mathbf{k})\hat{c}^{\dagger}_{\mathbf{k},l+1}\hat{c}_{\mathbf{k},l}-\Gamma^{\dagger}(\mathbf{k})\hat{c}^{\dagger}_{\mathbf{k},l}\hat{c}_{\mathbf{k},l+1}\right).
    \end{aligned}
\end{equation}
To simplify the expression, we assume the layers are equally spaced: $z_{l+1}-z_{l}=d$ for all $l$. We obtain
\begin{equation}\label{}
    \begin{aligned}[b]
        [\hat{r}_z,\hat{v}_z]=\frac{1}{i\hbar}[\hat{r}_z,\hat{H}(\mathbf{k})]=\frac{d^2}{i\hbar}\sum^{N-1}_{l}\left(\Gamma(\mathbf{k})\hat{c}^{\dagger}_{\mathbf{k},l+1}\hat{c}_{\mathbf{k},l}+\Gamma^{\dagger}(\mathbf{k})\hat{c}^{\dagger}_{\mathbf{k},l}\hat{c}_{\mathbf{k},l+1}\right).
    \end{aligned}
\end{equation}
Therefore, 
\begin{equation}\label{}
    \begin{aligned}[b]
        X_{zz}=\frac{-d^2}{\hbar^2}\sum^{N-1}_{l}\sum_{n\mathbf{k}}\mel{u_{n\mathbf{k}}}{\left[\Gamma(\mathbf{k})\hat{c}^{\dagger}_{\mathbf{k},l+1}\hat{c}_{\mathbf{k},l}+\Gamma^{\dagger}(\mathbf{k})\hat{c}^{\dagger}_{\mathbf{k},l}\hat{c}_{\mathbf{k},l+1}\right]}{u_{n\mathbf{k}}}f_{n\mathbf{k}}.
    \end{aligned}
\end{equation}
Furthermore, the off-diagonal term $X_{za}(a=x,y)$ is given by 
\begin{equation}\label{}
    \begin{aligned}[b]
        X_{za}=-\frac{1}{\hbar^2}\sum_{n\mathbf{k}}f_{n\mathbf{k}}\langle u_{n\mathbf{k}}|[\hat{r}_z,[\hat{r}_a,\hat{H}(\mathbf{k})]|u_{n\mathbf{k}}\rangle.
    \end{aligned}
\end{equation}
We have $[\hat{r}_a,\hat{H}(\mathbf{k})]=i\frac{\partial \hat{H}(\mathbf{k})}{\partial \mathbf{k}_a}$, and $[\hat{r}_z,[\hat{r}_a,\hat{H}(\mathbf{k})]]$ gives rise to the term which only contains the interlayer hopping. We thus find
\begin{equation}\label{}
    \begin{aligned}[b]
        X_{za}=&-\frac{d}{\hbar^2}\sum_{n\mathbf{k}}\sum^{N-1}_{l}f_{n\mathbf{k}}\mel{u_{n\mathbf{k}}}{\frac{\partial \Gamma(\mathbf{k})}{\partial k_a}\hat{c}^{\dagger}_{\mathbf{k},l+1}\hat{c}_{\mathbf{k},l}-\frac{\partial \Gamma^{\dagger}(\mathbf{k})}{\partial k_a}\hat{c}^{\dagger}_{\mathbf{k},l}\hat{c}_{\mathbf{k},l+1}}{u_{n\mathbf{k}}}\\
        =&-\frac{d}{\hbar^2}\sum_{n\mathbf{k}}f_{n\mathbf{k}}\mel{u_{n\mathbf{k}}}{\frac{\partial \hat{J}_{-}(\mathbf{k})}{\partial k_a}}{u_{n\mathbf{k}}},
    \end{aligned}
\end{equation}
where $\hat{J}_{-}(\mathbf{k})=\sum^{N-1}_{l}\Gamma(\mathbf{k})\hat{c}^{\dagger}_{\mathbf{k},l+1}\hat{c}_{\mathbf{k},l}-\Gamma^{\dagger}(\mathbf{k})\hat{c}^{\dagger}_{\mathbf{k},l}\hat{c}_{\mathbf{k},l+1}$.

For the special case of a bilayer, the expressions can be simplified by introducing Pauli matrices $\tau_0,\tau_x,\tau_y,\tau_z$ in the two-dimensional layer space.  This leads us to a Hamiltonian of the form 
\begin{equation}
    \begin{aligned}[b]
        \hat H_{BL}(\mathbf{k})=h(\mathbf{k})\tau_0+\Gamma_1(\mathbf{k})\tau_{+}+\Gamma^{\dagger}_1(\mathbf{k})\tau_{-}+V \tau_z,
    \end{aligned}
\end{equation}
where $h(\mathbf{k})$ is the monolayer  Hamiltonian (a matrix in sublattice space) , $\Gamma_1$ is the  interlayer hopping matrix, and $V$ is the interlayer displacement field. Here we used the standard symbols $\tau_{+}=\frac{1}{2}(\tau_x+i\tau_y)$ and $\tau_{-}=\frac{1}{2}(\tau_x-i\tau_y)$. The $z-$components of the position and velocity operators are then
\begin{equation}
    \begin{aligned}[b]
        \hat{r}_z=&\frac{d}{2}I\tau_z\\
        \hat{v}_z=&\frac{1}{i\hbar}[\hat{r}_z, H_{BL}]=\frac{d}{i\hbar}\left(\Gamma_1\tau_{+}-\Gamma^{\dagger}_1\tau_{-}\right),
    \end{aligned}
\end{equation}
where $I$ is the identity matrix in sublattice space, and $d$ is the spacing between layers. We  can thus calculate  $X_{zz}$ as
\begin{equation}\label{X_zz}
    \begin{aligned}[b]   
        X_{zz}=-\frac{d^2}{\hbar^2}\left\langle\left(\Gamma_1\tau^{+}+\Gamma^{\dagger}_1\tau^{-}\right)\right\rangle_F\,,
    \end{aligned}
\end{equation}
where Fermi sea average $\langle...\rangle_F$ is defined in Eq.~(\ref{X_tensor}) of the main text.

\section{The derivation of the in-plane components  $X_{ab}$}

We present our own derivation of Eq.~(\ref{OurDrude}) of the previous section. The tensor $X_{ab}$ defined in Eq.~(\ref{X_tensor}) can be rewritten as:
\begin{equation}\label{def}
    \begin{aligned}[b]
         X_{ab}=-\frac{1}{\hbar^2}  \left\langle[\hat{r}_a,[\hat{r}_b,\hat{H}_0]]\right\rangle_F\,.
    \end{aligned}
\end{equation}
We can immediately verify that it is symmetric under the interchange of \( a \) and \( b \):
\begin{equation}\label{}
    \begin{aligned}[b]
         X_{ab}=-\frac{1}{\hbar^2}\left\langle[\hat{r}_b,[\hat{r}_a,\hat{H}_0]]\right\rangle_F+\frac{1}{\hbar^2}\left\langle[\hat{H}_0,[\hat{r}_a,\hat{r}_b]]\right\rangle_F=X_{ba}\,,
    \end{aligned}
\end{equation}
using the Jacobi identity \( [\hat{A},[\hat{B},\hat{C}]] + [\hat{B},[\hat{C},\hat{A}]] + [\hat{C},[\hat{A},\hat{B}]] = 0 \).
Next, we split the position operator \(\hat{\mathbf{r}}\) into its intraband and interband components, denoted by \(\hat{\mathbf{R}}\) and \(\hat{\mathbf{X}}\) respectively: \(\hat{\mathbf{r}} = \hat{\mathbf{R}} + \hat{\mathbf{X}}\). The well-known representations are:
\begin{equation}\label{rep}
    \begin{aligned}[b]
        [\hat{\mathbf{R}}]_{n\mathbf{k},n'\mathbf{k}'} &= \left( i\partial_{\mathbf{k}} \delta_{\mathbf{k},\mathbf{k}'} + i\langle u_{n\mathbf{k}}|\partial_{\mathbf{k}} u_{n\mathbf{k}}\rangle \delta_{\mathbf{k},\mathbf{k}'} \right) \delta_{n,n'} \\
        [\hat{\mathbf{X}}]_{n\mathbf{k},n'\mathbf{k}'} &= i\langle u_{n\mathbf{k}}|\partial_{\mathbf{k}} u_{n'\mathbf{k}}\rangle (1 - \delta_{n,n'})
    \end{aligned}
\end{equation}
Inserting above expressions into the Eq.~\eqref{def}, we have
\begin{equation}\label{sep}
    \begin{aligned}[b]
         \left\langle[\hat{r}_a,[\hat{r}_b,\hat{H}_0]]\right\rangle_F=&\left\langle\hat{R}_a\hat{R}_b\hat{H}_0+\hat{H}_0\hat{R}_b\hat{R}_a-\hat{R}_a\hat{H}_0\hat{R}_b-\hat{R}_b\hat{H}_0\hat{R}_a\right\rangle_F\\
         +&\left\langle\hat{X}_a\hat{X}_b\hat{H}_0+\hat{H}_0\hat{X}_b\hat{X}_a-\hat{X}_a\hat{H}_0\hat{X}_b-\hat{X}_b\hat{H}_0\hat{X}_a\right\rangle_F,
    \end{aligned}
\end{equation}
where $\hat{H}_0$ is the $\mathbf{k}$-dependent. Further, we have
\begin{equation}\label{intra}
    \begin{aligned}[b]
    \left\langle\hat{R}_a\hat{R}_b\hat{H}_0+\hat{H}_0\hat{R}_b\hat{R}_a-\hat{R}_a\hat{H}_0\hat{R}_b-\hat{R}_b\hat{H}_0\hat{R}_a\right\rangle_F=\left\langle(\hat{R}_a\hat{R}_b\hat{H}_0)\right\rangle_F,
    \end{aligned}
\end{equation}
where $(\hat{R}_a\hat{R}_b\hat{H}_0)$ means the operator only acts on the $\hat{H}_0$, and this terms can be simplified as $-\frac{\partial ^2\hat{H}_0}{\partial k_a\partial k_b}$. The interband parts in Eq.~\eqref{sep} have a simple form by using Eq.~\eqref{rep} 
\begin{equation}\label{inter}
    \begin{aligned}[b]
    \left\langle\hat{X}_a\hat{X}_b\hat{H}_0+\hat{H}_0\hat{X}_b\hat{X}_a-\hat{X}_a\hat{H}_0\hat{X}_b-\hat{X}_b\hat{H}_0\hat{X}_a\right\rangle_{n\mathbf{k}}=&2\Re\sum_{m,m\neq n}\left(\epsilon_{n\mathbf{k}}-\epsilon_{m\mathbf{k}}\right)[\hat{X}_a]_{nm}[\hat{X}_b]_{mn},
    \end{aligned}
\end{equation}
Here we have employed 
\begin{equation}
    \begin{aligned}[b]
        \mel{\partial_a u_{n\mathbf{k}}}{\hat{H}_0-\epsilon_{n\mathbf{k}}}{\partial_b u_{n\mathbf{k}}}=&\sum_{m,m\neq n}\braket{\partial_a u_{n\mathbf{k}}}{u_{m\mathbf{k}}}\mel{u_{m\mathbf{k}}}{\hat{H}_0-\epsilon_{n\mathbf{k}}}{u_{m\mathbf{k}}}\braket{u_{m\mathbf{k}}}{\partial_b u_{n\mathbf{k}}}\\
        =&\sum_{m,m\neq n}(\epsilon_{m\mathbf{k}}-\epsilon_{n\mathbf{k}})[\hat{X}_a]_{nm}[\hat{X}_b]_{mn}.
    \end{aligned}
\end{equation}
Combining Eq.~\eqref{intra} and Eq.~\eqref{inter}, we arrive at the final expression of effective inverse mass tensor
\begin{equation}\label{def2}
    \begin{aligned}[b]
         X_{ab}=&\sum_{n\mathbf{k}}\left[\frac{1}{\hbar^2} \left\langle\frac{\partial ^2\hat{H}_0}{\partial k_a\partial k_b}\right\rangle_{n\mathbf{k}}-2\Re\mel{\partial_a u_{n\mathbf{k}}}{\hat{H}_0-\epsilon_{n\mathbf{k}}}{\partial_b u_{n\mathbf{k}}}\right]f_{n\mathbf{k}}\\
         =&\frac{1}{\hbar^2}\sum_{n\mathbf{k}}\frac{\partial^2 \epsilon_n(\mathbf{k})}{\partial k_a \partial k_b}f_{n\mathbf{k}}.
    \end{aligned}
\end{equation}

Alternatively, we can write:
\begin{equation}\label{}
    \begin{aligned}[b]
        [\hat{r}_a,[\hat{r}_b,\hat{H}_0]] = \hat{r}_a\hat{r}_b\hat{H}_0 + \hat{H}_0\hat{r}_b\hat{r}_a - \hat{r}_a\hat{H}_0\hat{r}_b - \hat{r}_b\hat{H}_0\hat{r}_a\,.
    \end{aligned}
\end{equation}
In the space of parameters \(\mathbf{k}\) of the band-\(n\) manifold, the position operator \(\hat{r}_a\) has the form:
\begin{equation}\label{}
    \begin{aligned}[b]
        \hat{r}_a = i\frac{\partial}{\partial k_a} - A^a_{nn}(\mathbf{k})\,.
    \end{aligned}
\end{equation}
It is straightforward to verify that:
\begin{equation}\label{}
    \begin{aligned}[b]
        X_{ab} = \frac{1}{\hbar}\sum_{n\mathbf{k}} f_{n\mathbf{k}} \frac{\partial v^b_{nn}(\mathbf{k})}{\partial k_a} = \frac{1}{\hbar^2}\sum_{n\mathbf{k}} f_{n\mathbf{k}} \frac{\partial^2 \epsilon_n(\mathbf{k})}{\partial k_a \partial k_b}\,.
    \end{aligned}
\end{equation}
Thus, \( X_{ab} \) is the average of the averaged effective inverse mass tensor:
\begin{equation}\label{}
    \begin{aligned}[b]
        \left[ \frac{1}{m^*} \right]_{ab} = \frac{1}{\hbar^2} \frac{\partial^2 \epsilon_n(\mathbf{k})}{\partial k_a \partial k_b}\,.
    \end{aligned}
\end{equation}
The expression of tensor $X$ in the periodic directions is exactly the same as the Drude weight except for a constant $2\pi e^2$.

\section{Symmetry analysis of the $X_{ab}$ tensor}\label{}
In this section, we will give an analysis of the symmetry constraints of $X_{ab}$ based on BLG model. In the unbiased state, Bernal bilayer graphene belongs to the \( D_{3d} \) point group, characterized by a combination of inversion, rotation, and reflection symmetries. However, the application of a perpendicular electric field introduces an asymmetry between the two graphene layers, reducing the symmetry to the \( \hat{c}_{3v} \) point group. This lower symmetry includes a three-fold rotation axis (C3) and three vertical mirror planes, but lacks inversion symmetry. 

1. Three-fold rotation axis $(C3)$: This involves rotating the bilayer by 120° about an axis perpendicular to the graphene planes.

2. Mirror planes $(\mathcal{M}_{ab})$: There are three mirror planes that contain the three-fold rotation axis and bisect the angles between the nearest-neighbor carbon atoms in the same plane.

We consider about three-fold rotation $\hat{c}_3$ symmetry along $z$ direction ($\hat{c}_{3z}$), three mirrors are parallel to $z$ direction and three $\mathcal{M}$ symmetries. We can write down the symmetry transformation matrices
\begin{equation}\label{}
    \begin{aligned}[b] 
        \hat{c}_{3z}=
        \begin{pmatrix}
            \cos\frac{2\pi}{3} & -\sin\frac{2\pi}{3} & 0\\
            \sin\frac{2\pi}{3} & \cos\frac{2\pi}{3} & 0 \\
            0 & 0 & 1
        \end{pmatrix},\quad
        \mathcal{M}_{xz}=
        \begin{pmatrix}
            1 & 0 &0\\
            0 & -1 & 0 \\
            0 & 0 & 1
        \end{pmatrix}.
    \end{aligned}
\end{equation}
Under $\hat{c}_{3z}$ transformation, we have:
\begin{equation}\label{}
    \begin{aligned}[b]
        (\hat{c}_{3z}X \hat{c}^{-1}_{3z})=\begin{pmatrix}
            \frac{1}{4}(X_{xx}+3X_{yy})+\frac{\sqrt{3}}{4}(X_{xy}+X_{yx}) & \frac{1}{4}(X_{xy}-3X_{yx})+\frac{\sqrt{3}}{4}(X_{yy}-X_{xx}) & -\frac{X_{xz}}{2}-\frac{\sqrt{3}}{2}X_{yz}\\
            \frac{1}{4}(X_{yx}-3X_{xy})+\frac{\sqrt{3}}{4}(X_{yy}-X_{xx}) & \frac{1}{4}(3X_{xx}+X_{yy})-\frac{\sqrt{3}}{4}(X_{xy}+X_{yx}) & \frac{\sqrt{3}}{2}X_{xz}-\frac{X_{yz}}{2} \\
            -\frac{X_{zx}}{2}-\frac{\sqrt{3}}{2}X_{zy} & \frac{\sqrt{3}}{2}X_{zx}-\frac{X_{zy}}{2} & X_{zz}
        \end{pmatrix}
    \end{aligned}
\end{equation}
If $\hat{c}_{3z}$ symmetry is preserved, we have $X_{xx}=X_{yy}$, $X_{xy}=-X_{yx}$, and $X_{xz}=X_{yz}=X_{zx}=X_{zy}=0$.
Under $\mathcal{M}_{xz}$ transformation, we have:
\begin{equation}\label{}
    \begin{aligned}[b]
        (\mathcal{M}_{xz}X \mathcal{M}^{-1}_{xz})=\begin{pmatrix}
            X_{xx} & -X_{xy} & X_{xz}\\
            -X_{yx} & X_{yy} & -X_{yz}\\
            X_{zx}&-X_{zy} & X_{zz}
        \end{pmatrix}
    \end{aligned}
\end{equation}
If $\mathcal{M}_{xz}$ symmetry is preserved, we have $X_{xy}=X_{yx}=X_{zy}=X_{yz}=0$. Combining rotational symmetry and mirror symmetry, we can verify that the off-diagonal terms of the $X$ tensor vanish.

\section{Calculation of $X_{zz}$ in Bernal-Stacked Bilayer Graphene}
The $4\times4$ tight-binding (TB) model for Bernal-stacked bilayer graphene band is~\cite{McCann2013} 
\begin{equation}
    \begin{aligned}[b]
        H=\sum_{l={1,2},ij}\gamma_0e^{-i\phi^{l,l}_{i,j}}\hat{c}^{\dagger}_{li}\hat{c}_{lj}+\sum_{ij}\gamma_1e^{-i\phi^{1,2}_{i,j}}\hat{c}^{\dagger}_{1i}\hat{c}_{2j}
        +\sum_{i'j'}\gamma_3e^{-i\phi^{1,2}_{i',j'}}\hat{c}^{\dagger}_{1i'}\hat{c}_{2j'}+h.c.+\sum_{l={1,2},ij}V^{l}\hat{c}^{\dagger}_{li}\hat{c}_{li},
    \end{aligned}
\end{equation}
where $\gamma_0$ is the intralayer nearest neighbor hopping constant, $\gamma_1$ is the interlayer direct hopping, and $\gamma_3$ is the interlayer hopping between A and B sublattices. $V$ is the potential difference between the layers, created by a vertical electric field. With a non-zero bias $V$ the gap is located at the two inequivalent $K$ and $K'$ points of the Brillouin zone (BZ). 
We neglect the terms $\gamma_3$ and $\gamma_4$ and only present the minimal TB model. The Hamiltonian reduces to
\begin{equation}
    \begin{aligned}[b]
        H_{0}(\mathbf{k})=-\gamma_0\mathbf{h}(\mathbf{k})\cdot\bm{\sigma}\tau_0+\frac{\gamma_1}{2}\left(\sigma_x\tau_x+\sigma_y\tau_y\right)+V\sigma_0\tau_z.
    \end{aligned}
\end{equation}
The velocity operators have the following forms
\begin{equation}
    \begin{aligned}[b]
        \hat{v}_x=&-\gamma_0\left(\partial_{k_x}F(\mathbf{k})\sigma_x+\partial_{k_x}G(\mathbf{k})\sigma_y\right)\tau_0,\\
        \hat{v}_y=&-\gamma_0\left(\partial_{k_y}F(\mathbf{k})\sigma_x+\partial_{k_y}G(\mathbf{k})\sigma_y\right)\tau_0,\\
        \hat{v}_z=&\frac{\gamma_1d}{2\hbar}\left(\sigma_x\tau_y-\sigma_y\tau_x\right).
    \end{aligned}
\end{equation}
The $z$-component of the position operator is written as $\frac{d}{2}\sigma_0\tau_z$. We  calculate  $X_{zz}$, which is given by 
\begin{equation}
    \begin{aligned}[b]   
    X_{zz}=&\frac{d}{2i\hbar}\langle[\sigma_0\tau_z,\hat{v}_z]\rangle_F
    =-\frac{\gamma_1 d^2}{2\hbar^2}\langle\left(\sigma_x\tau_x+\sigma_y\tau_y\right)\rangle_F.
    \end{aligned}
\end{equation}

\end{document}